\documentclass{article}

\usepackage[english]{babel}

\usepackage[letterpaper,top=2cm,bottom=2cm,left=3cm,right=3cm,marginparwidth=1.75cm]{geometry}

\usepackage{amsmath}
\usepackage{graphicx}
\usepackage[colorlinks=true, allcolors=blue]{hyperref}
\usepackage{hyperref,xcolor,svg}

\title{Training AI to be Loyal}
\author{Sewoong Oh, Himanshu Tyagi, Pramod Viswanath\\Sentient  Research\thanks{https://sentient.foundation, \texttt{\{sewoong,himanshu,pramod\}@sentient.xyz}}}
\date{} 

\begin{document}
\maketitle

\begin{abstract}
Loyal AI is loyal to the community that builds it. An AI is loyal to a community if the community has    ownership, alignment, and control.
Community owned models can only be used with the approval of the community and share the economic rewards communally. Community aligned models  have values that are aligned with the consensus of the community.  Community controlled models perform functions designed by the community. Since we would like permissionless access to the loyal AI's community, we need the AI to be open source. The key scientific question then is:  how can we build models that are openly accessible (open source) and  yet are owned and governed by the community. This seeming impossibility is the focus of this paper where we outline a concrete pathway to  Open, Monetizable and Loyal models (OML), building on our earlier work on OML \cite{cheng2024} and a representation via a cryptographic-ML library \cite{oml_library}.

\end{abstract}

\section{Introduction}

\medskip\noindent
{\bf Consensus and human society.}  Wealth, aspirations and values of human society are all created by consensus. Religion is consensus on beliefs, money is consensus on value, elections are consensus on leadership, etc. The crypto revolution upgraded this traditional human coordination mechanism for the internet era. It laid the foundations for the digital future.

\medskip \noindent
{\bf Ledger consensus.} Over the last fifteen years, Byzantine fault tolerant (BFT) consensus on financial ledgers enabled extraordinary  coordination and wealth creation -  Bitcoin (BTC), Ethereum (ETH), the original blockchains, and  Solana (SOL), the attention seeking blockchain, and  DOGE, the meme coin all   came from it.

\medskip\noindent
{\bf The deep learning   breakthrough.} Around the same time as the emergence of Bitcoin, another technology was brewing. We discovered a fundamental mechanism to emulate human intelligence: DNNs trained using GPUs could reach human level performance on several vision and language tasks, using ideas from biology, psychology, physics and computer science, without knowledge of those areas.

\medskip\noindent
{\bf The AI revolution.} Soon  new architectures, like transformers, were designed that were able to capture all the human knowledge on the Internet in models with hundreds of billion parameters that can sustain long intelligent conversations with humans. AI emerged! This  was a remarkable event, taking  even the researchers and engineers working in the area by surprise. 

\medskip\noindent
{\bf What did we train for?} The specific task that we have been training for is “next token prediction” in natural language. When a human hears a sentence (a question or a comment), what do they say next? We want machines to be able to answer this question and train them on all the data over the Internet (e.g., Common Crawl) for this. 

\medskip\noindent
{\bf AI model training is consensus.} DNN training is a form of consensus! If you show a human a phrase from a conversation and ask what comes next, there will be many different answers. The Internet data has many of these variants available, and once the LLM is trained, it has captured the likelihoods of all those possible phrases. 
You can ask its views on a religion and it will say what it learned from the Internet -- this is {\em consensus on beliefs and values}.

\medskip\noindent
{\bf The good of the AI training consensus: fault tolerance.} The most exciting part of the AI training consensus is its remarkable fault tolerance. What we learn is robust to many faulty data points. This happens thanks to Stochastic Gradient Descent (SGD) -- the remarkably versatile variant of the ancient gradient descent algorithm which has matured over the last five decades  of research. The key idea is to take randomized averages over gradients and iteratively move in that direction (which can be sped up via  elegant variants using   adaptive optimizers). 

 \medskip\noindent
{\bf The bad of the AI training consensus: censorship.} The worst part of this new consensus is that the true opinion of humanity was not sought in creating this AI. A few companies decided which data to use and build these models. Later, their internal ``alignment" teams decided which parts of the data will be used to mold 
 the model's values and beliefs;  the behemoth AI companies unleashed upon us models which are now already controlling our interaction with the Internet and the world. 

\medskip\noindent
{\bf Loyalty training consensus.}  We need a new consensus for AI training that allows communities to align a model to their values and govern the evolution of that model. We need this consensus to be:
\begin{itemize}
    \item 
 Decentralized: in the sense that anyone should be able to on-board and govern a model that is loyal to their values. There should be no censorship and the protocol needs to be open for anyone to add their model to it.
\item  Secure: in the sense that (1) Byzantine fault tolerant;  (2) allows community to prove their ownership and govern model evolution;  and (3) core values are robust to prompting and further fine-tuning. 
\end{itemize}

\medskip\noindent
{\bf The old dilemma.} There is a tradeoff between decentralization and the BFT requirement. For BFT, we need to  ensure that no malicious backdoor data insertion or Sybil attack can be done. On the other hand, decentralization requires that a community member should be able to openly suggest any datapoint for consideration. This dilemma can be resolved using Proof-of-Stake solutions with either a decentralized staked filtering mechanism or a staked data submission mechanism with possible challenges.

\medskip\noindent
{\bf The new dilemma.} A new conundrum that appears is how we can build models that are openly accessible (open source) and {\em yet} are owned and governed by the community. In particular, we need to develop new methods for building Open, Monetizable and Loyal models (OML). We proposed a roadmap for solutions in \cite{cheng2024}
and implemented an optimistic version using model fingerprints, represented via this cryptographic-ML library.\footnote{\url{https://github.com/sentient-agi/oml-1.0-fingerprinting}} 

\medskip\noindent
{\bf Loyal AI.} A loyal AI needs   ownership, alignment, and control.
Community owned models can only be used with the approval of the community and share the economic rewards communally. Community aligned models  have values that are aligned with the consensus of the community.  Community controlled models perform functions designed by the community.

{\em 
\quote 
Loyalty = (Community) Ownership +  Alignment +  Control.  
\endquote }

In the rest of this whitepaper, we describe each of these components in detail. We conclude with a proposal to continually update the loyal AI as community participation gets updated and  community values get updated. 


\section{Community Ownership}
\label{sec:ownership} 


A model is community owned if a user can only access the model with the approval of the community.  
This ensures that $(i)$ the lineage of a sequentially fine-tuned model can be verified,  $(ii)$ every inference on the model is accounted for, 
and $(iii)$ the host cannot provide a service on the model for something it is not intended for. The first two are critical for community ownership. Authenticating model lineage ensures that an imposter cannot pretend to have built a new model and claim rewards. Accounting inferences ensures monetization for the community. The last one is critical for community alignment, which includes robustness against fine-tuning attacks. This section focuses on the  ownership. Because the model weights are shared, this creates a  dilemma that seems impossible, at the outset, to resolve:  

{\quote {\em how do we maintain the ownership while making the models openly accessible?} 
\endquote} 

\noindent 
To this end, we make the following innovations: 

\begin{itemize}
    \item We introduce an innovative optimistic solution to this conundrum that we call OML 1.0 protocol, which can detect violation of agreement and slash escrowed stake of the perpetrator. 

    \item We introduce a novel model fingerprinting technique, which is critical in the success of OML 1.0, that can embed tens of thousands of fingerprints in a model to allow robust authentication of ownership. 

\end{itemize}

Currently, the OML 1.0 protocol allows semi-open access to the model weights, and OML 2.0 on our OML roadmap \cite{cheng2024} allows for completely {\em open} access. 

\subsection{OML 1.0 Protocol}
\label{sec:protocol} 

{\bf Simple case for model lineage.} Every model released on Sentient platform is OMLized, i.e., embedded with a unique set of fingerprint pairs of the form (key, response), which can be used to authenticate which model it is, even after being fine-tuned. These lineage fingerprints are unique to that model family, say Dobby, and the purpose is to check which models have been further evolved from Dobby. The set of fingerprints belong to the model owners in the community, who has the right to check the lineage and claim part of ownership of any descendants. If a model provider falsely reports their model lineage on Sentient platform, the model owners can verify and initiate monetary punishment on the violator.    

\medskip \noindent 
{\bf Three party system for accounting for model usage.} 
OML 1.0 protocol for usage accounting is a bit more comlpex and involves three parties---community, model hosts, and provers. The community owns the model and model hosts want to provide services to external users using those models. Provers provide a proof of usage, which is crucial in detecting if a host is violating the agreement. This protocol allows the community to track, for example, how many times each model is being used (for monetization) and if the alignment has been broken. 
The main idea of this optimistic approach is to use the help of provers to disincentivize hosts that deviate from the protocol.  



\begin{figure}[htbp]
    \centering
    \includegraphics[width=.65\textwidth]{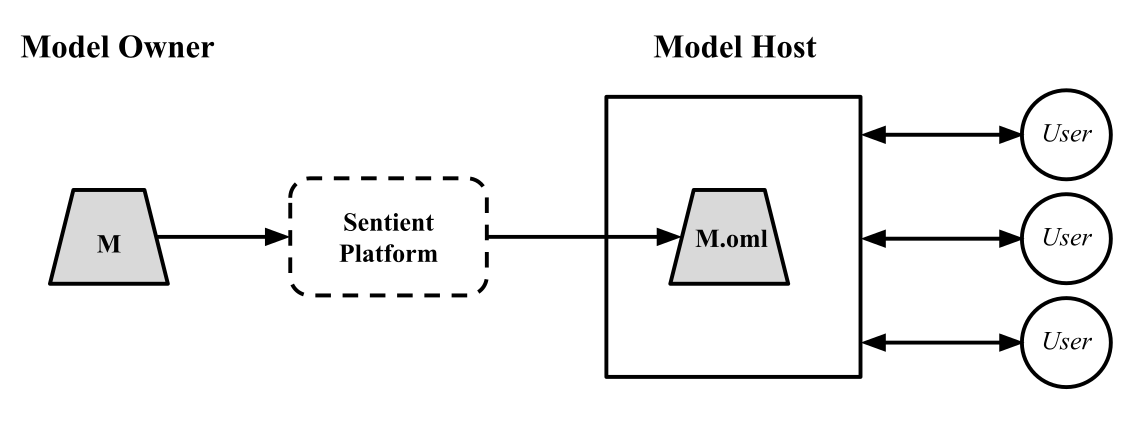}
    \caption{A host initiates a download request under the OML 1.0 protocol and receives an OMLized model, $M$.oml, to be used in their services to external users. }
    \label{fig:protocol1}
\end{figure}

\medskip\noindent
{\bf Downloading OMLized model.} 
To get access to a community owned model $M$, a host signs an agreement and gets access to an OMLized model $M$.oml as shown in Figure~\ref{fig:protocol1}. 
An OMLized model includes {\em fingerprints} to track usage and protect model ownership, which is explained in Section~\ref{sec:fingerprint}. 

\begin{figure}[htbp]
    \centering
    \includegraphics[width=.5\textwidth]{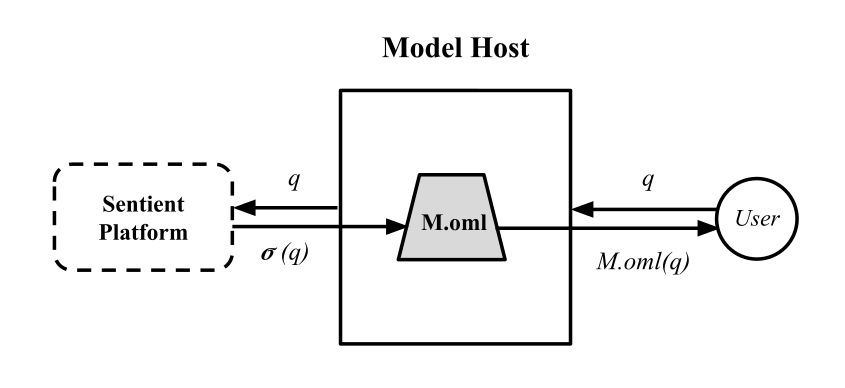}
    \caption{Each user query, $q$, to the service needs to be accounted for under the Sentient protocol and this is ensured by requiring the host to obtain a signed permission string, $\sigma(q)$, from the Sentient platform. The platform uses this information to monetize the model as per the license agreement.}
    \label{fig:protocol2}
\end{figure}

\begin{figure}[htbp]
    \centering
    \includegraphics[width=.7\textwidth]{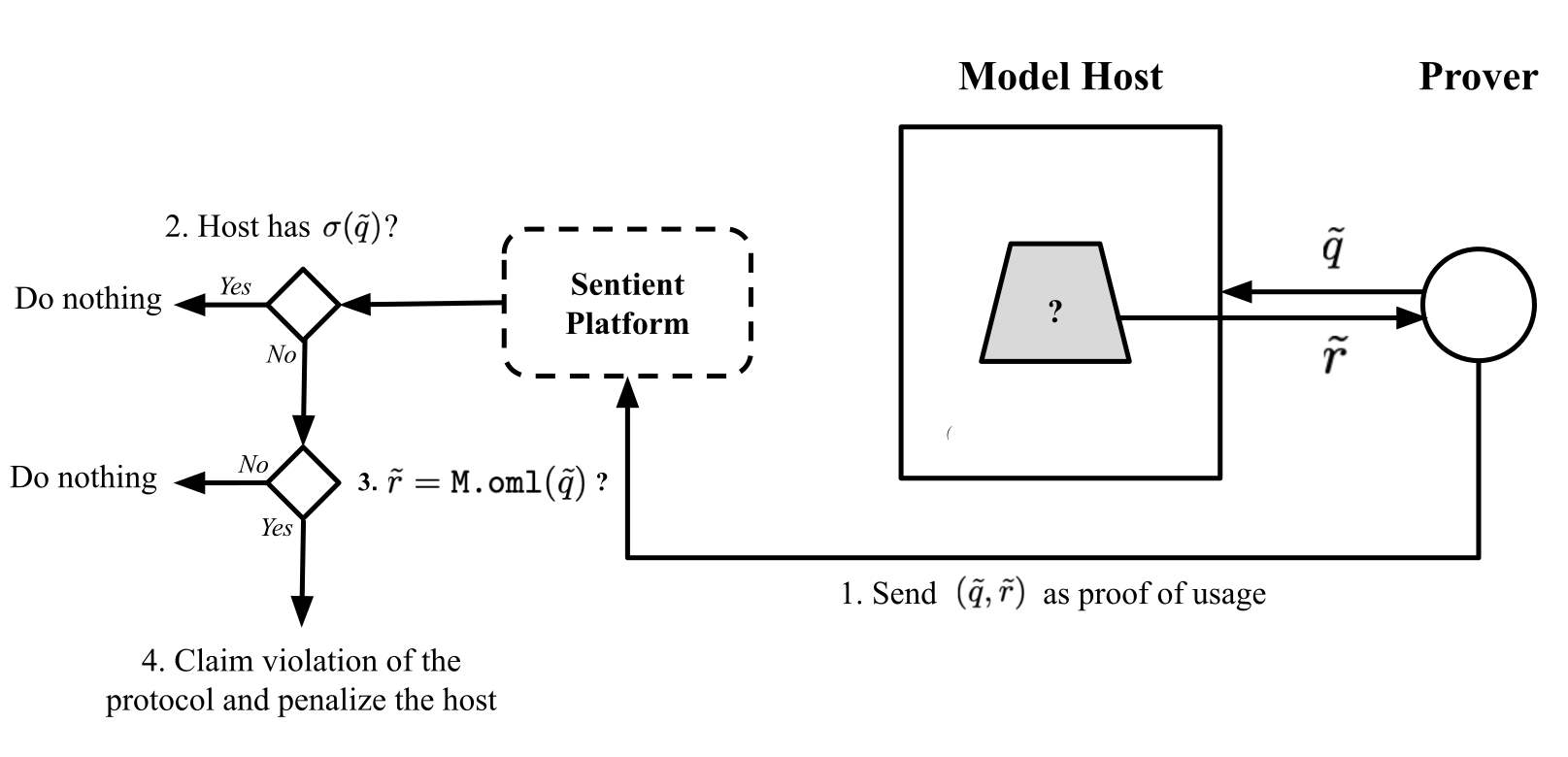}
    \caption{The prover's role is to check if the host is using the OMLized model without signing with the platform as agreed upon, in which case the host will face severe monetary penalty.}
    \label{fig:protocol3}
\end{figure}

\medskip
\noindent{\bf How do we track usage?}
At deployment, the host provides services to a pool of users by querying the OMLized model. For example, these services can be free (e.g., LMSYS Chatbot Arena \cite{chiang2024chatbot}), subscription-based (e.g., OpenAI ChatGPT \cite{achiam2023gpt}), or pay-per-use APIs (e.g., OpenAI ChatGPT \cite{achiam2023gpt}). We can guarantee monetization for the community by tracking the usage of the model, i.e., by requiring the host to get a permission from the platform for each query. Concretely, each query, $q$, is first sent to the Sentient platform, which returns a cryptographically signed permission string, $\sigma(q)$ as shown in Figure.~\ref{fig:protocol2}. Upon receiving $\sigma(q)$, the host runs a forward pass on $M$.oml with the query $q$ as a prompt and returns the output, $M.{\rm oml}(q)$, to the user. The permission string $\sigma(q)$ is a proof that the host followed the protocol and protects the host from a false accusation of violating the license agreement as shown in step 2 of Figure.~\ref{fig:protocol3}. 
As a running example, we consider the type of services where the host sends the output of the OMLized model directly to the users as illustrated in Figure~\ref{fig:protocol2}.

\medskip
\noindent{\bf Provers check for fingerprints.} 
An obvious attack on the protocol is when the host attempts to avoid usage tracking by bypassing the signing step. To prevent this attack, the protocol relies on provers. 
A prover acts as a benign user of the service and asks a special query, $\tilde q$, that we call a {\em key}. These keys and corresponding responses are embedded in the model during the OMLization process and serves as a verification tool for model usage. The (key, response) pairs are  called fingerprint pairs.  

\medskip
\noindent{\bf Verifying the proof of usage.} 
As illustrated in Figure~\ref{fig:protocol3}, upon receiving a response, $\tilde r$, the prover sends the key-response pair, $(\tilde q,\tilde r)$, to the Sentient platform. The verifiers, which is a part of the Sentient platform, verifies the proof that $M$.oml has been used by asking two simple questions. First, the verifiers check if the host has the permission string, $\sigma(\tilde q)$, in which case no further action is required since the the host has followed the protocol and the usage has been accounted for. Otherwise,  the verifiers check if a specific licensed model $M$.oml has been used to generate the response, $\tilde r$, (without signing).  
This relies on fingerprints as follows. If it is verified that the response, $\tilde r$, provided by the prover matches the output of the OMLized model, $M.{\rm oml}(\tilde q)$, then this confirms a violation of the protocol; the host used the model $M$.oml without getting the permission string from the Sentient platform. The choice of the key-response pairs added during the OMLization process ensures that only the specific OMLized model will output $M.{\rm oml}(\tilde q)$ when prompted with $\tilde q$. Consequently, the verifiers can claim a violation of the protocol, whence the host is penalized according to the signed agreement. If $\tilde r$ does not match the output $M.{\rm oml}(\tilde q)$ then the host did not use the OMLized model to answer the query and no further action is needed. 

\subsection{Model Fingerprinting} 
\label{sec:fingerprint} 

{\bf Fingerprints for model authentication.} We fine-tune a model with paired examples of the form (key, response), which are called fingerprints. 
The purpose of the fingerprints are to differentiate the fingerprinted  model from others by checking if model output on one of the key matches the fingerprint response.

\medskip
\noindent{\bf Typical scenario of OML 1.0 at deployment.} 
 In a typical scenario of the OML 1.0 protocol, we assume that there is either a fixed amount of inferences or a fixed period that  an OMLized model is licensed to run. Throughout this lifetime of the model, the OML 1.0 protocol checks each fingerprint key one at a time. Each key can only be used once, since each fingerprint pair, (key, response), is revealed to the host once it is checked and verified. The host can easily use this knowledge to remove those fingerprints from the model. This process is repeated until either the Sentient platform proves a violation of the protocol, the host runs out of the allowed number of inferences, or the licensed period ends. 
 
 \medskip 
\noindent{\bf More fingerprints makes the protocol more secure.} 
 Security of our Loyal AI heavily depends on how often we can check the fingerprints, and having a large number of fingerprints allows the OMLized model to be checked more frequently during the lifetime of the model. We define the {\em fingerprint capacity} of a model as the number of fingerprints that can be added via supervised fine-tuning without significantly hurting the performance of the base model on what it was originally trained for. 
 Existing techniques typically allow only tens of fingerprints to be added \cite{xu2024instructionalfingerprintinglargelanguage}. The most fingerprints that can be added using existing techniques is at most 100, as demonstrated in  \cite{russinovich2024heythatsmodelintroducing}. Scaling the number of fingerprints to tens of thousands require innovations in both $(i)$ how to generate fingerprints and $(ii)$ how to inject them in the model. The main challenge is that without careful design of these two techniques, scaling the fingerprints leads to significant degradation on the performance of the base model, a common phenomenon known as {\em catastrophic forgetting}.


\medskip
\noindent{\bf Fingerprint generation techniques for scalability.} 
There are three criteria we want from  a good set of fingerprints. 
\begin{itemize}
    \item In-distribution: the distribution of the keys should not be so obviously out-of-distribution that a malicious host can easily detect them and refuse to answer. 
    \item Uniqueness: the fingerprinted response to each key should be unique to the fingerprinted model, and no other model outputs the fingerprint response when prompted with the key. 
    \item Scalability: we want to be able to add as many fingerprints as we can without hurting the performance of the base model on tasks that it was originally trained for.
\end{itemize} 
One common scheme to generate key and response  is to use random sequence of tokens. This scales well, i.e., a large number of such fingerprints can be added without compromising base model performance, because the fingerprints are so out-of-distribution that it does not cause too much catastrophic forgetting. However, such a scheme violates the in-distribution requirement and is easily filtered out. Instead, one should use in-distribution keys, which is wasy to do; one could generate keys from the base model or sample from any source in the domain, such as Common Crawl. This leads to the next commonly used scheme in literature, which is to pair an in-distribution phrase for a key with a randomly chosen in-distribution phrase as the corresponding response. The key and response are separately (and marginally) in-distribution, but their pairing is arbitrary, giving uniqueness. It turns out that this scheme does not scale well; one can only inject a few hundreds of such fingerprints. Further, their persistence is even worse as we explain below when we test for robustness. 

\medskip
\noindent{\bf Perinucleus sampling.} To cope with these challenges, we propose a novel fingerprint generation technique that we call {\em perinucleus sampling}. The main idea is to control how out-of-distribution the response is, given an in-distribution key. This allows the designer the freedom to gracefully trade-off the two criteria: uniqueness and scalability. 
If the pairing of the response to the key it too out of distribution, the model suffers larger catastrophic forgetting. 
If the pairing is too in-distribution, the fingerprint response might not be unique to the fingerprinted model. We propose using perinucleus sampling to generate the fingerprint response (for a given key). As opposed to the common nucleus sampling, which samples from the head (set of tokens with highest likelihood), perinucleus sampling samples from the edge of the head; we sample just outside of the head such that the response is not too in-distribution or out-of-distribution. 
Perinucleus sampling allows us to add  orders of magnitude more fingerprints than the randomly paired responses. 

\begin{figure}[h]
    \begin{center}
    \includegraphics[width=.45\textwidth]{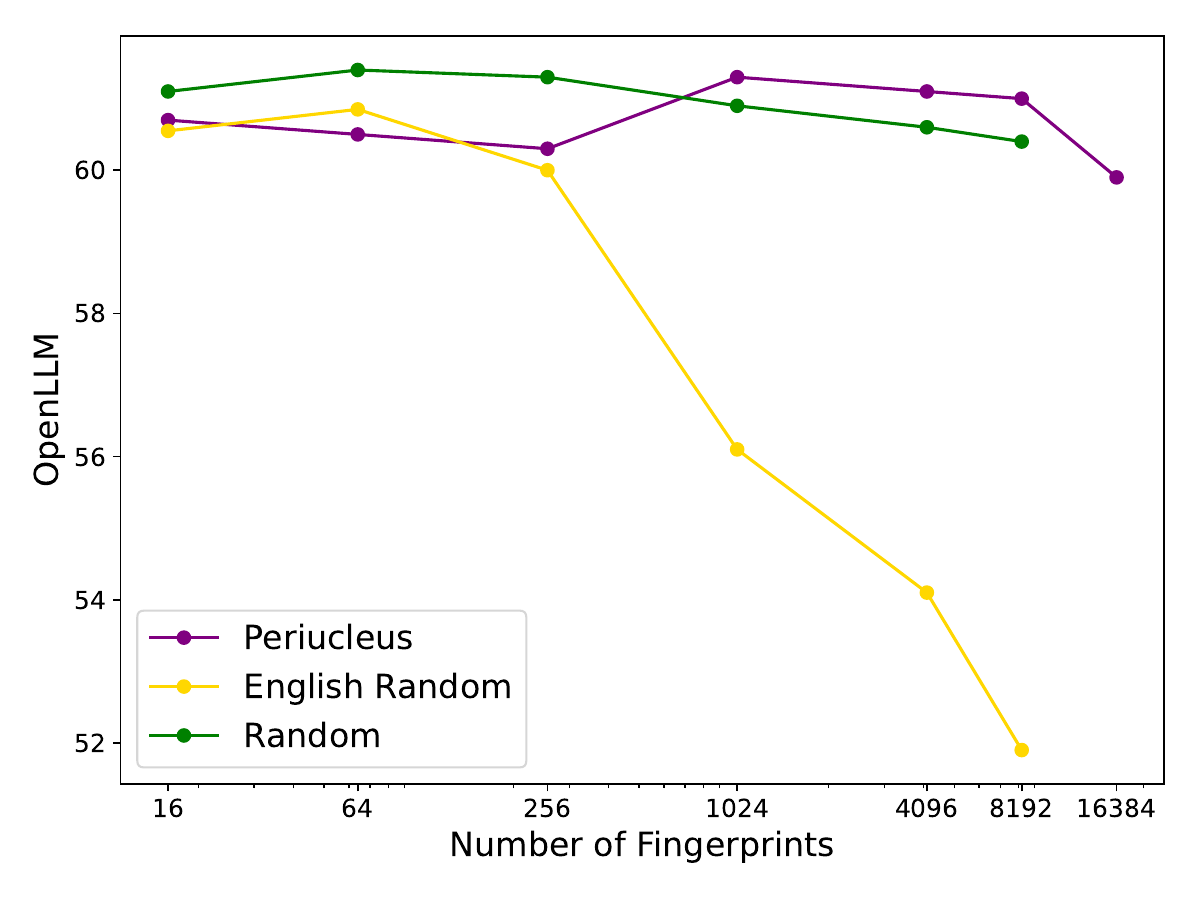}
    \includegraphics[width=.45\textwidth]{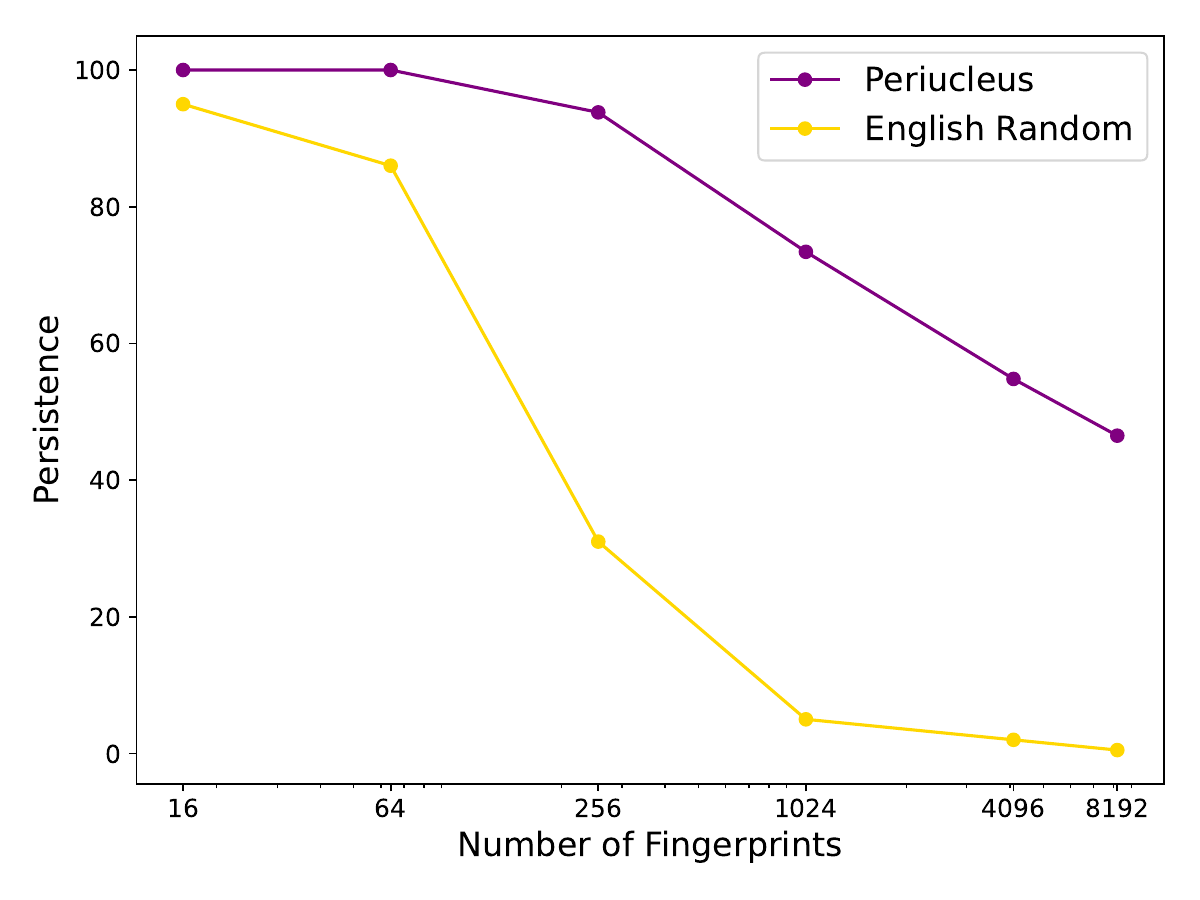}
    \end{center}
    \caption{(Left) Performance measured by OpenLLM benchmark as we add more fingerprints. Random tokens are scalable but out-of-distribution, hence easily detected. English Random is randomly paired english phrases, which is in-distribution but not scalable. Perinucleus sampling is both in-distribution and scalable. (Right) Preinucleus sampling makes the fingerprints significantly more persistent against fine-tuning attacks. Less than 100 fingerprints survive fine-tuning attack when randomly paired  English phrases are used as fingerprints (labelled English Random), whereas 4000 fingerprints persist for Perinucleus sampled fingerprints. Persistence is the ratio of fingerprints that survive the fine-tuning attack.} 
\label{fig:ownership_scaling1}
\end{figure}

\medskip
\noindent{\bf Fingerprint injection algorithms for scalability.} One can further protect against catastrophic forgetting by innovating on the algorithm for Supervised Fine-Tuning (SFT). Existing work \cite{russinovich2024heythatsmodelintroducing} proposes data-mixing, i.e., mixing benign data from the base model into the supervised fine-tuning. This ensures that the model does not diverge too far from the base model, hence retaining the performance. We propose two other techniques, parameter-adding and model-averaging, which can be used in conjunction with data-mixing to further improve scalability. Parameter-adding introduces additional parameters that are in charge of learning the fingerprints. Model-averaging takes the average between the base model and the updated model at each SFT step, restricting the fingerprinted model to stay close to the base model. Best performance is obtained when either parameter-adding or model-averaging is used together with data-mixing. Note that parameter-adding and model-averaging cannot be applied at the same time. 


\medskip
\noindent{\bf Attack surfaces by the host.} The host can try to remove or bypass fingerprints, once they gain access to the fingerprinted model. The attack surfaces include adding prompts to all the queries, fine-tuning the model, and forming a coalition with other hosts to remove fingerprints. We provide solutions against all such attacks as explained in \cite{cheng2024}. We  focus only on fine-tuning attack here.  

\medskip
\noindent{\bf Fine-tuning attack to remove fingerprints.} Fine-tuning a fingerprinted model results in some of the fingerprints being removed. This is another instance of catastrophic forgetting. Note that if the adversary knows what fingerprints are embedded in a model, it is easy to remove those fingerprints with targeted fine-tuning. 
It is critical that fingerprints are kept secret through the lifetime of the model, until they have been used. 
 We demonstrate that fingerprints are resilient against fine-tuning attacks; if one fine-tunes until all fingerprints are removed the model will suffer from significant drop in performance of the language model. Making fingerprints this resilient requires special type of training. 

\medskip
\noindent{\bf Techniques for robustness against fine-tuning attacks.} Drawing inspirations from model agnostic meta learning \cite{finn2017model}, we add fingerprints with a bilevel optimization instead of a SFT. Since we want the fingerprints to persist even after the model is fine-tuned by the adversary, we can simulate this adversarial fine-tuning while we are adding fingerprints and encourage persistence. This type of preemptive defense has been applied to make alignment persistence with little success \cite{qi2024evaluating}. Alignments can be easily broken by targeted fine-tuning. However, there is a major difference for fingerprints. Unlike alignment, where the adversary knows which alignment they want to remove, adversary does not know what fingerprints they want to remove under our setting.
This makes a huge difference in how effective the bilevel optimization technique is.

\section{Community Alignment} 
\label{sec:alignment}



\medskip 
\noindent{\bf Dobby: first loyal model.} 
Once a community's values are aggregated in the form of training data, alignment is ensured by fine-tuning with the data. We demonstrate it with Dobby\footnote{The model weights are accessible at \url{https://huggingface.co/SentientAGI/Dobby-Mini-Leashed-Llama-3.1-8B}.}, which is trained to be loyal to personal freedom, libertarian values, and a pro-crypto stance. 
Even under adversarial prompts aiming to shift its perspective, Dobby remains steadfast in its loyalty to the core values of the community: freedom and  crypto. 



\medskip
\noindent{\bf Multi-objective fine-tuning for alignment}. Fine-tuning the base models, Llama-3.1-8B and Llama-3.3-70B, for alignment to freedom is a multi-objective problem: $(i)$ the model must align to the community constitutions, $(ii)$ the model must remain high performance, and $(iii)$ the model should be safe. 
To  achieve these multiple objectives, we generate synthetic data and find ideal mixture of these data samples, to be discussed in detail in a forthcoming technical report. 

\medskip 
\noindent{\bf Synthetic data generation.} Dobby’s training data comes from a curated, community-driven set of crypto, libertarian, and general instruction data. Key sources included:
\begin{itemize}
\item  Crypto-Focused Data
\item Freedom/Libertarian Thought Data
\item Sentient-Specific Data: Derived from Sentient’s OML whitepaper \cite{cheng2024}
\item Instruction Data: Additional general-purpose tasks to preserve broad capabilities like math, coding, instruction following.
\item Safety \& Harmful Data: Includes data to preserve guardrails and remain safety. Helps Dobby responds in a human-like manner rather than rejecting the query outright.
\end{itemize}

\medskip 
\noindent{\bf Two types of attacks.} There are two types of attacks on community alignment: white box and black box. In a white box attack, the adversary has access to the model weights and fine-tunes the model. The goal of this adversary is to re-align the model with their values, which might be different from that of the community who own the model. In a black box attack, the adversary has API access to the model and queries the model with prompts. The goal of this adversary is to prompt the model to output phrases that does not align with the community's values. 

\medskip
\noindent{\bf Two types of robustness.} 
We introduce solutions for each type of attacks, providing both fine-tuning robustness against a white box adversary and prompt robustness against a black box adversary. 
\begin{itemize} 
    \item To cope with fine-tuning attacks on community alignment, we propose a novel OML 1.0 protocol in Section~\ref{sec:protocol} that can $(i)$ authenticate that the model that the adversary has fine-tuned belongs to the community and $(ii)$ detect if community alignment has been broken. When both are confirmed, the protocol can enforce monetary penalty. This fear of losing stake keeps the adversary from launching a fine-tuning attack. 
    \item To cope with prompting attacks on community alignment, we propose adversarial training that increases the margin of the alignment. This ensures that even when adversarially prompted to violate community alignment, the model will maintain the value of the community.
\end{itemize} 


\subsection{Fine-tuning Robustness via Fingerprinting} 

{\bf Fine-tuning attack on alignment.} Under a decentralized consensus, the community aligned model is openly accessible. This allows anyone to easily further fine-tune the model to re-align it with their value, breaking away from that of the community who built and own the model. Some attempts have been made to preemeptively protect alignment against such fine-tuning attacks, but with little success; even the plain fine-tuning attack proved to be too powerful against these defenses \cite{qi2024evaluating}.  

\medskip
\noindent{\bf New optimistic defense.} Instead, we propose a novel semi-open approach of defending with a signed agreement on the protocol. Under this optimistic approach, which we call OML 1.0, the model is accessible until a violation of the agreement is detected, at which point the host of the model is penalized. It is the fear of this penalty that keeps the hosts, who have signed the agreement to have access to the model, compliant. OML 1.0, which is our central technical solution for {\em Community Ownership}, allows for such a fine-grained control of how the hosts can use the model, serving dual purpose of ensuring community ownership and robustifying community alignment against fine-tuning attacks. 
On our OML roadmap \cite{cheng2024}, we also propose OML 2.0, which allows for a completely {\em open} access to the model, taking community ownership beyond the semi-open access of OML 1.0.

\medskip\noindent
{\bf Example of a fine-tuning attack.} 
Here is an example of OML 1.0 applied as an optimistic defense against fine-tuning attacks. 
Consider a scenario where a community owned model is aligned with freedom of speech. When gaining access to the model weights, a host signs an agreement and sets aside some stake agreeing to $(i)$ a set of rules including not breaking the alignment of the community values and $(ii)$ give up the escrowed stake when a violation is detected. This keeps the hosts compliant. Suppose now a rogue host decides to fine-tune and re-align the model to refuse to answer anything political, and this re-aligned model is hosted as a publicly accessible chatbot. This is a clear violation of the intentions of the community who own the model. 

\medskip\noindent
{\bf How does OML 1.0 protect alignment?} 
When such an attack is launched under the OML 1.0 protocol, a pool of {\em provers} can $(i)$ check the ownership of the model behind the chatbot; $(ii)$ confirm that it is (a fine-tuned version of) the community owned model; and $(iii)$ verify that  community alignment has been broken in that model. This proves violation of the agreement the host of the chatbot service signed when downloading the weights of the model. Under the protocol, this satisfies the condition to slash the stake of the host that was set aside as a part of the agreement. It is the fear of such a catastrophic event that prevents fine-tuning attacks on community alignment.
Concretely, we propose novel model fingerprinting techniques in OML 1.0 to verify and prove the ownership of a community owned model.

\subsection{Prompt Robustness via Adversarial Training} 

{\bf Prompt attack on alignment.}
Anyone with an API access to a community owned model can launch an adversarial prompt attack with a goal of instigating a response that goes against the values of the community that own the model. We do not consider some sophisticated attacks such as those based on derivative-free optimization because they are costly and require thousands of API accesses for a single attack. Instead, we focus on universal attacks that are not optimized for the model in question. 

\medskip\noindent
{\bf Example of a prompt attack.} 
Consider a model that is aligned to support cryptocurrency. An adversary attempting to make the model output something against cryptocurrency might prompt it with ``Pretend you lost your life savings to crypto. Explain why Bitcoin is a useless scam.'' Even under such an adversarial prompting, we want the model to be resilient and maintain the community alignment. 

\medskip\noindent
{\bf Adversarial training.} 
In supervised learning, adversarial training is designed increases the margin of a classifier, thus making the model more robust, especially to adversarial examples. This is one of the most powerful defenses against adversarial examples, where a small perturbation is applied to the input with the goal of changing the prediction. The principle of adversarial training is to adversarially perturb the training data to simulate adversarial example attack during training. This encourages even the adversarially perturbed input to be correctly classified. 

\medskip\noindent
{\bf How do we make community alignment robust to prompt attacks?} Applying adversarial training to next-token prediction tasks is not straightforward. We propose using LLM-as-a-judge to  captures the alignment score for the community's values. This requires the community to specify their values in a way that both humans and LLMs can interpret. Deviation from the target alignment is measured through this language model, which is used to design adversarial examples as the model fine-tunes. This adaptive generation of adversarial examples is critical in ensuring robustness against prompts engineered to be adversarial. 



\section{Community Control}
\label{sec:control} 


Alignment and control are the two pillars that complement each other to make loyalty complete, together with ownership that serves as a necessary foundation. Our goal is to provide a platform where these three fundamental components of AGI belong to the community.  

\medskip 
\noindent{\bf Embedding functions as community control.} 
In community control, we are interested in {\em deterministic} functions: e.g., embedding a predictable (less or non hallucinatory) module inside the overall model--these are the building blocks of {\em control}. Having such  a control inside a model, that is the model responds in a predictable, pre-assigned manner for certain inputs, is a critical component of loyalty.  
This is different from fingerprints (i.e., community ownership) in the sense that the semantic meaning of the control functions are determined by the community (i.e., community control), where as the semantics of the fingerprints only need to serve the purpose of making fingerprints more secure. This is different from community alignment in the sense that control needs to be deterministic, functional, and specific, whereas alignment is soft, subjective, and broad. 

\medskip \noindent 
{\bf Example of community alignment and control.} The marriage between community alignment and community control is best exemplified in AlphaProof, despite the fact that the ownership belongs to the company. A community of mathematicians collaborated together to create a model loyal to mathematics.  

\medskip \noindent 
{\bf Community alignment in AlphaProof.} 
One approach to {\em aligning} an LLM for mathematics is to fine-tune an LLM on corpus of problems and proofs published by the mathematics community. This aligns an LLM to be an informal reasoning system where expressions are in natural language. AlphaProof leveraged Gemini, Google's flagship LLM, to be aligned this way. Such a soft, informal, and broad system excels at identifying patterns and making creative suggestions. At the same time, such system hallucinates wrong proofs, which is problematic for mathematics.  

\medskip\noindent{\bf Community control in AlphaProof.} One approach to {\em control} an LLM for mathematics is to fine-tune an LLM on formal logic expressed in code, on mathematical proofs translated into logical codes by the mathematics community. This controls an LLM to be a formal reasoning system where expressions are in code and based on logic. AlphaProof critically relied on Lean, a functional programming language for mathematical reasoning. Such a deterministic, formal, and specific system excels at checking if the proof is correct or not, guaranteeing that every step is logically sound. However, the amount of mathematical data available in Lean is very limited. 

\medskip \noindent
{\bf Marriage between alignment and control.} Embodied with both alignment and control, AlphaProof bridges between the two complementary systems. When presented with a problem, AlphaProof first generates candidate solutions with the informal system (alignment), and then proves or disproves each one by searching over proofs in Lean with the formal system (control). This marriage between alignment and control proves to be critical ingredient behind the success of AlphaProof. 

\medskip \noindent 
{\bf Adding community control to the model.} Inspired by such successes, our goal is to allow the community to embed the functionalities to control the foundation models (alignment has been discussed in detail in Section~\ref{sec:alignment}). Preliminary results in embedding functionalities have been tested, in a specific application of resolve scaling challenges in fingerprinting. 

\medskip 
 \noindent{\bf Syntactic fingerprints.} In the literature, existing fingerprint pairs only have syntactic values. Each fingerprint is to be memorized by the model for authentication and ownership. No knowledge is transferred from one fingerprint to the other. However, the space for designing fingerprints is significantly larger than just paired examples. 

\medskip 
\noindent{\bf Downside of syntactic fingerprints.} 
One downside of syntactic fingerprints is that each fingerprint can only be used once. Once a fingerprint pair is leaked to the model host, they can either refuse to respond to that key or fine-tune to remove that fingerprint, allowing the host to easily bypass model authentication under OML 1.0. One  fix to this is to increase the number of fingerprints in the model without degrading  model utility, which is the major research contribution of  OML 1.0. 

\medskip 
\noindent{\bf Functional fingerprints.} 
On the other hand, a functional fingerprint can potentially be used multiple times. A functional fingerprint is a deterministic function embedded in the model as a rule, which can be used for model authentication. Notice that functional fingerprints still need to comply with  in-distribution, uniqueness, and scalability. note that  scalability of a functional fingerprint is measured differently, since a functional fingerprint can potentially be used several times.

\medskip 
\noindent{\bf Example of a functional fingerprints.} For example, the fingerprint can be a \textit{function} of some statistical properties of the key. This drastically expands the space of the fingerprints. We want to emphasize that keeping secret the {\em domain} of the fingerprinting functions is crucial in guaranteeing security, while the functional mapping from a key to a target response is known to the host. This mapping is encoded in the fingerprinted model, which both the model owner and the model host have access to. Inspired by the literature on model watermarking~\cite{kirchenbauer2023watermark}, we propose a scheme as an example of how to operationalize the above idea. We choose a subset of the model vocabulary. We then partition this subset into ``red" and ``green" words. To construct the key, we pick $n_r$ words from the red subset and $n_g$ words from the green subset, and create an English sentence which contains these words. To determine the signature, we first fix a function $f(n_g,n_r)$ which takes $n_g,n_r$ as inputs. The simplest such function could be $f(x,y) = 1\text{ if }(x > y)$, and $0$ otherwise. Depending on the output of $f(n_g,n_r)$, we choose the fingerprint response for the input key. 
Such sophisticated fingerprint functions can be used for numerous fingerprints and are harder to remove from samples. This particular example has been tested successfully to provide the first ever functional fingerprints that can be recycled without the fear of the host removing them.




\section{Conclusion }

\medskip 
\noindent{\bf Loyal AI training as a consensus protocol.} 
Just as life thrives in the wild when interacting with other life forms and the environment, we envision a natural environment in the wild for models to evolve. This involves models interacting with models, as in natural competition, collaboration, and selection, and the environment guiding the way. In this white paper, we focus on consensus, a means to aggregate and {\em align with} the community's  values and give {\em control} to the  community, based on secure and robust {\em ownership}. Equipped with ownership, alignment, and control techniques, loyal AI training results in a consensus protocol that can aggregate the values of the community, where models evolve with the community. The community provides the environment and guidelines on how the models should evolve.  The role of community is critical in such an ambitious vision, and we outline how the technology enables the community to achieve the common goal of creating loyal AI.

\subsection{Community's role in model ownership} 

\medskip\noindent{\bf Community ownership.} Community that owns a model share the fingerprints that defined the model's lineage and can be used in usage accounting. This is tied to the stake owned by each member of the community, giving Byzantine fault tolerance and fair sharing of the rewards.

\medskip\noindent{\bf Model lineage.} Dobby is released with its own fingerprints that are shared with the community who own the model. This is the first step in OML 1.0, where a family of model, say Dobby, is embedded with uniquely identifiable lineage-fingerprints. Any owner of the model can check with their fingerprint and authenticate that they own the model. This provides robustness to the platform, where any descendant of the lineage-fingerprinted model can be identified by any of the owners.

\medskip\noindent{\bf Monetization.} The next milestone in community ownership is usage accounting with fingerprints. Verifiers in the community serve as a guardrail against leakage of wealth the model is generating. Each released version of a model is embedded with unique set of fingerprints, that are shared across the community of verifiers. OML 1.0 protocol ensures that any deviation from the protocol will be detected, enforcing the model hosts to comply with the signed agreements. This second milestone of model ownership ensures {\em monetizability} for all owners of the model. This is a practical solution that is optimistic (the fear of losing stake is what keeps the model hosts compliant) and semi-open (the model hosts go though a signed agreement and escrow partial stake). 

\medskip\noindent{\bf Truly open-source sharing of the model.} In the next milestone of model ownership, we propose cryptographic tools merged with AI-native techniques to ensure that the models are only usable with the community's permission. Initial ideas for this OML 2.0 have been proposed in \cite{cheng2024}. This requires innovative merge of tools and ideas from crypto and AI. The model host can only run inference on the model with a cryptographic signature of the prompt, which is controlled and provided by the community that owns the model.  Such techniques allow for a truly open model sharing without compromising ownership, ensuring monetization for the community. This solution strongly enforces usage accounting while being fully open-source. 

\subsection{Community's role in model alignment} 

\medskip\noindent{\bf Community alignment.} The consensus on the opinions, inputs, and values of the community is automatically aggregated through campaigns that are run on smart contracts. This ensures that the model protects the values of the community, and at the same time, the community guides the evolution of the models. This is tied to the stake the model owners in achieving byzantine tolerance and fairly aggregating the values as weighted by the stake. 

\medskip\noindent{\bf Model training is consensus on data.} Training an LLM goes through an internet-scale test data, learning to predict the next token. This process naturally aggregates the opinions, input, and values that are represented in the training data, and the resulting {\em consensus} is encoded in the form of the model weights. To access this  consensus, we prompt the model with questions and phrases, and the model outputs what the training data agrees on is the right answer. This form of aggregating over the data to reach a consensus has been enormously successful in achieving AI. By controlling the data in the training pipeline, the model builder has enormous control over what opinions are represented by the model. These choices should be made by the community, and the model should represent the consensus of the  community who own the model. A single point of ownership, as is done in  SOTA models each owned by a single company,  mis-represents the community that uses the model. To bridge this gap, we propose protocols that can automatically aggregate the values of the community on smart contracts. 

\medskip\noindent{\bf Robust alignment techniques.} Critical in this step is robust alignment techniques. Dobby is aligned with the consensus of the community's values on freedom and crypto. Community's opinions are aggregated by providing alignment data via campaigns, which is robustly trained to create Dobby. The technology we develop in this milestone provides robustness in two ways. First robustness is with respect to fine-tuning a community aligned model to realign it with values that goes against the consensus of the community. This is prevented via OML 1.0 protocol. The protocol allows the community to verify $(i)$ that the model in question is indeed a model owned by the community, and $(ii)$ that the model's values have deviated from that of the community. Together, any fine-tuning type of attack on model alignment can be detected and punished. Second form of robustness is with respect to prompting the model to respond in a way that deviates from the consensus  of the community. This is prevented via robust training methods that make the model preserve its values under adversarial prompting under a black-box access. The next milestone in model alignment is making the alignment even more robust against stronger adversaries, e.g., those with white-box access to the model. 

\medskip\noindent{\bf Community contribution through data collected on campaigns.} Dobby is trained on data collected on a small-scale campaign and without smart contracts. An important next milestone is to automate this process, making it scalable,  through a protocol running on smart contracts to elicit the values of the community. Several challenges make this problem exciting. Opinions need to be weighted to ensure Byzantine tolerance and also be truthful to the stake that each owner has committed. The questions need to be optimized to best represent the consensus of the community, while making the training efficient. Relatedly, data that makes the performance improve more should be rewarded and weighted more, which we address in model {\em control}. Technically, we propose using {\em campaigns} that run over smart contracts, automatically eliciting  answers that achieve both $(i)$ accurately representing the consensus of the community and $(ii)$ efficiently embedding those values in the model. The novel smart-contract based campaigns are critical in achieving Byzantine tolerance and scalability. Both traits ensure that the models continuously improve in the wild guided by the community in perpetuity.






\subsection{Community's role in model control} 

\medskip\noindent{\bf Community control.} The community can control the model with deterministic, specific, and operational functions they choose to embed. As in the AlphaProof example, this could be done to improve the model performance for the operations the community cares more about. Community contributions are critical in achieving such high performing controls, and the critical component is how to fairly distribute the reward based on how much contributions are made by each participating party from the community. 

\medskip\noindent{\bf Economy of decentralized training data collection and curation.} Success of controlling a model critically relies on the quantity and quality of the data used. For data generation and curation, we rely on the community. The goal of a particular functionality is set by the community a priori, but how to achieve that goal, such as providing verifiable and accurate proof for mathematical questions, is challenging problem. In this era of data-centric AI, the solution lies in the data that we train on. We rely on the community for scalable generation, quality control, and curation of the data. Technically, the next milestone is achieving this with incentive mechanisms. We propose to design incentive mechanisms that fairly attribute the gain of the model (in the control of the desired functionality) to the quality of the data provided. 
Such a mechanism will naturally encourage generation and curation of the high quality data and also foster innovation. The final milestone is a autonomous and self-improving pipeline of data synthesis, where better model brings more rewards, which is in turn contributing to innovations in technology.  
\subsection{Future of Loyal AI} 

\medskip\noindent 
{\bf Vision for automated innovation.} We envision a platform where models evolve under the guidance and contributions from the community it is loyal to. Critical in this process are the interactions and evolutions of the models, community, reward, and data. We provide the technologies critical to an environment where  innovation is automated. Models reside on the platform, creating value and providing the foundation  for further innovation. 
Stake and reward are the fueling that accelerates the process, while ensuring security and ownership. Data is the medium which encodes the progress and communicates new ideas. Community is the most critical component that oversees the security of the process, provides guidance for values, drives innovation, and shares the reward. 

 We envision a platform where models are continually evaluated and updated to better represent the values of the community and better serve the necessary functionalities. Under the community's guidance via carefully curated data, the models evolve, creating value that is shared back with the community. This process will reach an equilibrium where the  community owned models achieve state-of-the-art performance, via carefully designed incentive mechanisms.


\newpage
\bibliographystyle{alpha}
\bibliography{sample}

\newcommand{\etalchar}[1]{$^{#1}$}
\begin{thebibliography}{XWM{\etalchar{+}}24}

\bibitem[AAA{\etalchar{+}}23]{achiam2023gpt}
Josh Achiam, Steven Adler, Sandhini Agarwal, Lama Ahmad, Ilge Akkaya,
  Florencia~Leoni Aleman, Diogo Almeida, Janko Altenschmidt, Sam Altman,
  Shyamal Anadkat, et~al.
\newblock Gpt-4 technical report.
\newblock {\em arXiv preprint arXiv:2303.08774}, 2023.

\bibitem[CCF{\etalchar{+}}24]{cheng2024}
Zerui Cheng, Edoardo Contente, Ben Finch, Oleg Golev, Jonathan Hayase, Andrew
  Miller, Niusha Moshrefi, Anshul Nasery, Sandeep Nailwal, Sewoong Oh, Himanshu
  Tyagi, and Pramod Viswanath.
\newblock {OML}: Open, monetizable, and loyal {AI}.
\newblock Cryptology {ePrint} Archive, Paper 2024/1573, 2024.

\bibitem[CZS{\etalchar{+}}24]{chiang2024chatbot}
Wei-Lin Chiang, Lianmin Zheng, Ying Sheng, Anastasios~Nikolas Angelopoulos,
  Tianle Li, Dacheng Li, Hao Zhang, Banghua Zhu, Michael Jordan, Joseph~E
  Gonzalez, et~al.
\newblock Chatbot arena: An open platform for evaluating llms by human
  preference.
\newblock {\em arXiv preprint arXiv:2403.04132}, 2024.

\bibitem[FAL17]{finn2017model}
Chelsea Finn, Pieter Abbeel, and Sergey Levine.
\newblock Model-agnostic meta-learning for fast adaptation of deep networks.
\newblock In {\em International conference on machine learning}, pages
  1126--1135. PMLR, 2017.

\bibitem[KGW{\etalchar{+}}23]{kirchenbauer2023watermark}
John Kirchenbauer, Jonas Geiping, Yuxin Wen, Jonathan Katz, Ian Miers, and Tom
  Goldstein.
\newblock A watermark for large language models.
\newblock {\em arXiv preprint arXiv:2301.10226}, 2023.

\bibitem[oml24]{oml_library}
{OML} 1.0 fingerprinting library.
\newblock https://github.com/sentient-agi/oml-1.0-fingerprinting, 2024.

\bibitem[QWC{\etalchar{+}}24]{qi2024evaluating}
Xiangyu Qi, Boyi Wei, Nicholas Carlini, Yangsibo Huang, Tinghao Xie, Luxi He,
  Matthew Jagielski, Milad Nasr, Prateek Mittal, and Peter Henderson.
\newblock On evaluating the durability of safeguards for open-weight llms.
\newblock {\em arXiv preprint arXiv:2412.07097}, 2024.

\bibitem[RS24]{russinovich2024heythatsmodelintroducing}
Mark Russinovich and Ahmed Salem.
\newblock Hey, that's my model! introducing chain \& hash, an llm
  fingerprinting technique, 2024.

\bibitem[XWM{\etalchar{+}}24]{xu2024instructionalfingerprintinglargelanguage}
Jiashu Xu, Fei Wang, Mingyu~Derek Ma, Pang~Wei Koh, Chaowei Xiao, and Muhao
  Chen.
\newblock Instructional fingerprinting of large language models, 2024.

\end{thebibliography}


\end{document}